\documentclass[aps,prl,twocolumn,groupedaddress,amsmath,amssymb,showpacs]{revtex4}
\usepackage{graphicx}
\usepackage{color}
\usepackage{bm}
\usepackage{latexsym}

\def\a{\alpha}

\def\d{\delta}
\def\g{\gamma}

\def\om{\omega}

\newcommand{\ra}{\rangle}
\newcommand{\la}{\langle}

\def\RR{\mathbb{R}}

\def\CC{\mathbb{C}}

\newcommand{\cP}{\mathcal{P}}
\newcommand{\cT}{\mathcal{T}}
\newcommand{\cK}{\mathcal{K}}

\begin{document}

\title{ Bypassing the bandwidth theorem with $\cal PT$ symmetry}

\author{Hamidreza Ramezani$^1$, J. Schindler$^1$, F. M. Ellis$^1$, Uwe Guenther$^2$, Tsampikos Kottos$^{1}$}
\affiliation{$^1$Department of Physics, Wesleyan University, Middletown, CT-06459, USA}
\affiliation{$^2$Helmholtz Center Dresden-Rossendorf, POB 510119, D-01314 Dresden, Germany}

\date{\today}

\begin{abstract}
The beat time $\tau_{\rm fpt}$ associated with the energy transfer between two coupled oscillators is dictated by the bandwidth
theorem which sets a lower bound $\tau_{\rm fpt}\sim 1/\delta \omega$. We show, both experimentally and theoretically, that two
coupled active LRC electrical oscillators with parity-time (${\cal PT}$) symmetry, bypass the lower bound imposed by the bandwidth
theorem, reducing the beat time to zero while retaining a real valued spectrum and fixed eigenfrequency difference $\delta\omega$.
Our results foster new design strategies which lead to (stable) pseudo-unitary wave evolution, and
may allow for ultrafast computation, telecommunication, and signal processing.
\end{abstract}


\maketitle

One of the fundamental principles of wave physics is the Bandwidth Theorem \cite{G46} which in quantum mechanics
takes the form of the celebrated energy-time Heisenberg Uncertainty relation \cite{MT45}. A direct consequence
of this principle is the fact that the time for evolution  between two orthogonal states $\tau_{\rm fpt}$ (first
passage time) is bounded by $\tau_{\rm fpt}\sim 1/\delta \omega$ \cite{AA90,P93}. A basic example where this lower
bound can be exhibited is the beat time associated with the energy transfer between two coupled oscillators. Even
though the validity of the uncertainty principle is undoubted, it has recently
been suggested that possible extensions of quantum mechanics invoking non-Hermitian ${\cal PT}-$symmetric Hamiltonians \cite{BB98} can
generate arbitrarily fast state evolution referred to as brachistochrone dynamics \cite{BBJM07,note0,GS08b,BB09,note0-2}. The main
characteristic of this class of Hamiltonians ${\cal H}$ \cite{BBJM07,M07,GS08b,BB09} is that they commute
with an anti-linear operator ${\cal PT}$, where the time-reversal operator ${\cal T}$ is the anti-linear operator
of (generalized) complex conjugation and ${\cal P}$ is a (generalized) parity operator \cite{BB98}. Examples of such
${\cal PT}$-symmetric systems range from quantum field theories to solid state physics and classical optics \cite{BBJ02,
M02a,MGCM08,KGM08,BFKS09,GSDMVASC09,RKMCPS09,L09,K10,WKP10,S10,RKGC10,L10,FAHXLCFS11,SLZEK11,CGS11,LREKCC11,GGKN08}.
Due to the anti-linear nature of the ${\cal PT}$ operator,
the eigenstates of ${\cal H}$ may or may not be eigenstates of ${\cal PT}$. In the former case, all the eigenvalues of
${\cal H}$ are strictly real and the ${\cal PT}$-symmetry is said to be {\it exact}. Otherwise the symmetry is said to
be {\it spontaneously broken}. In many physical realizations, the transition from the exact to the broken ${\cal PT}$
-symmetric phase is due to the presence of various gain/loss mechanisms that are controlled by some parameter $\gamma$
of ${\cal H}$.

At the same time, the brachistochrone evolution has a long standing history and is significant both in theory and in application. It
was one of the earliest problems posed in the calculus of variations, and in the framework of classical mechanics
it dictates ``the curve down which a particle sliding from rest and accelerated by gravity will slip (without friction) from
one point to another in the least time'' \cite{brachi}. The quantum mechanical brachistochrone problem has recently been
revived in the emerging fields of quantum computation and signal processing where one studies the possibility to use
dynamical protocols in solving computational problems and enhancing signal transport respectively. More
specifically, the quantum brachistochrone problem can be formulated as follows: Given two orthogonal states $|\Psi_i
\rangle$ and $|\Psi_f\rangle$, one wants to {\it find} the (time-independent) Hamiltonian $H$ (protocol) that performs the
transformation $|\Psi_i\rangle\rightarrow |\Psi_f\rangle =e^{-iHt}|\Psi_i\rangle$ in the minimal time $\tau_{\rm fpt}$
for a fixed difference $\delta \omega=|(E_f-E_i)/\hbar|$ of the eigenvalues $E_f,E_i$ of $H$ \cite{CHKO06}. Such a constraint
is appropriate since a rescaling of the Hamiltonian as $H\rightarrow \lambda H$, with $\lambda>1$, would make $\delta
\omega$, and hence the transition rates, large. This corresponds to the fact that physically only a finite amount of
resources (e.g. a finite magnetic field, bandwidth resolution, etc.) are typically available. The quantum brachistochrone
was addressed by a number of researchers (see for example \cite{CHKO06,AA90,P93,L00,brody-03}) who show that the
minimal time to perform the required transformation is bounded by the energy-time uncertainty $\tau_{\rm fpt}\sim
{1\over \delta \omega}$.

Here we adopt these ideas into the realm of classical wave propagation and engineer a system comprised of two active $LRC$
circuits with ${\cal PT}$-symmetry having an arbitrarily low first passage time $\tau_{\rm fpt}$ bypassing the lower bound
imposed by the Bandwidth theorem. We call this limit-breaking wave phenomenon {\it tachistochrone passage} \cite{note} where
the equivalent system of coupled ${\cal PT}$-symmetric oscillators with the same bandwidth $\delta\omega$ can be obtained
if one oscillator experiences attenuation with rate $\gamma$ while its partner experiences an equivalent amplification rate
$\gamma$. Depending on the application point of the incident excitation, we observe unidirectional accelerated signal/energy
transport, where the time $\tau_{\rm fpt}\sim 1/\gamma$ can, in principle, become arbitrarily short. Although these results
rely on the fact that the generated dynamics is non-unitary, we mark that due to the ${\cal PT}$-symmetry, the corresponding
eigenfrequencies $\omega$ can be real, thus guaranteeing the stability of our system \cite{note2}.
Our
results foster new design strategies
based on active elements \cite{AR07} with ${\cal PT}$ symmetric arrangements that may allow for ultrafast computation,
telecommunication, and signal processing.


\begin{figure}
\centering
\includegraphics[clip,width=3.25in]{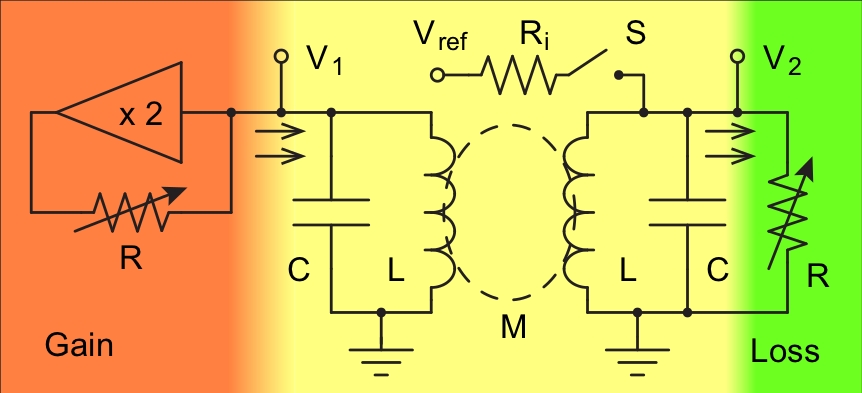}
\caption{
Electronic implementation of a ${\cal PT}$-symmetric dimer. The coils are inductively coupled with
$V_1$ and $V_2$ providing access to the system variables. The switch $S$ asserts the initial condition.
A resistor provides the damping (green side) while a negative resistance gain element (red side) is
implemented by feedback from an LM356 voltage doubling amplifier. Current flows in the direction of the arrows
proportional to the respective voltages $V_1$ and $V_2$.
}
\label{fig:fig2}
\end{figure}

Our system, shown in Fig.~\ref{fig:fig2}, consists of a pair of coupled $LC$ circuits, one with amplification and the
other with equivalent attenuation. The circuit was shown in Ref. \cite{SLZEK11} to be a simple realization of a
${\cal PT}$-symmetric dimer. Each inductor is wound with $75$ turns of \#28 copper wire
on $15 cm$ diameter PVC forms in a $6\times6 mm$ loose bundle for an inductance of $L = 2.32~mH$. The coils are
mounted coaxially with a bundle separation adjusted for the desired mutual inductance $M$.
The isolated natural frequency of each coil is $\omega_0=1/\sqrt{LC}=2\times10^5s^{-1}$.

The actual experimental circuit deviates from Fig.~\ref{fig:fig2} in the following
ways: (1) A resistive component associated with coil wire dissipation is compensated
by an equivalent gain component applied to each coil; (2) A small capacitance trim
is included to aid in circuit balancing; and (3) Additional LM356 voltage followers
are used to buffer the voltages $V_{1}$ and $V_{2}$, captured with a Tektronix
DPO2014 oscilloscope.

The linear nature of the system requires a balance of ${\cal PT}$ symmetry only to
the extent that component stability over time allows for a measurement. All circuit
modes either exponentially grow to the nonlinearity limit of the buffers, or shrink
to zero. Transient data is obtained respecting these time scales.

Kirchhoff's laws lead to the following set of equations for the charge $Q_1$ ($Q_2$) on the capacitor
corresponding to the amplified (lossy) side:
\begin{eqnarray}
 \label{kirchhoff2}
\frac{d^2Q_1}{d\tau^2} &=& -\alpha Q_1 + \mu \alpha Q_2 + \gamma \frac{dQ_1}{d\tau}\\
\frac{d^2Q_2}{d\tau^2} &=& \mu \alpha Q_1 - \alpha Q_2 - \gamma \frac{dQ_2}{d\tau}\nonumber
\end{eqnarray}
where $\tau\equiv \omega_0 t$, $\alpha=1/(1-\mu^2)\ge 1$, $\gamma=R^{-1}\sqrt{L/C}$ is the gain/loss parameter, and
$\mu = M/L$ is the rescaled mutual inductance. Inspection of Eqs. (\ref{kirchhoff2})
reveals that they are invariant under a combined parity (i.e. $Q_1\leftrightarrow Q_2$) and time-reversal (i.e. $t\rightarrow
-t$) transformation.

\begin{figure}
\belowcaptionskip=0pt
\includegraphics[clip,width=3in]{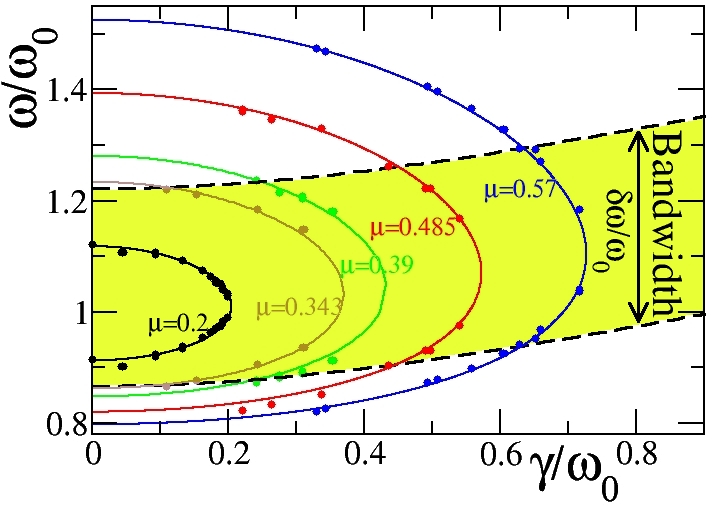}
\caption{
(Color online) Parametric evolution of the ${\cal PT}$-dimer eigenfrequencies vs. the gain/loss parameter $\gamma$, for
different mutual inductances $\mu$ (indicated in the figure). The solid circles are experimental data. The borders of the
highlighted area (marked by a black dashed line) indicate a constant bandwidth path with fixed frequency difference $\delta \omega/
\omega_0 =0.36$.
}
\label{fig:fig3}
\end{figure}


\begin{figure*}
\includegraphics[width=\hsize,keepaspectratio]{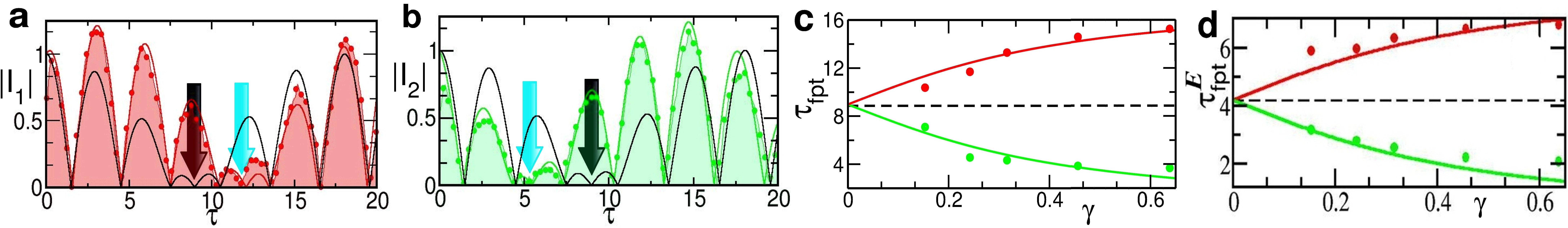}
\caption{(Color online) A representative tachistochrone passage wave propagation for a circuit with a bandwidth constraint $\delta\omega/
\omega_0=0.36$. (a) A typical temporal dynamics of the displacement current $I_1(\tau)$ when the initial condition corresponds
to an excitation of the circuit $n=1$ (gain side); (b) the same for current $I_2(\tau)$ when the initial condition is an
excitation at circuit $n=2$ (loss side). The black lines are numerical simulations for the passive $\gamma=0$ circuit, while
the red and green lines are numerical simulations for the active circuit with $\gamma=0.24$ and $\mu=0.39$. The measurements
are indicated with circles. The first passage time $\tau_{\rm fpt}$ is indicated with a cyan (black) arrow for the active
(passive) dimer. (c) The extracted $\tau_{\rm fpt}$ versus the gain/loss parameter $\gamma$. The green circles correspond to
an initial condition $I_2(0)=1$ (loss side) while the red circles to an initial condition
$I_1(0)=1$ (gain side). The solid lines indicate the theoretical result Eq.~(\ref{tautheory}). The black dashed
line indicates the beat time for a passive ($\gamma=0$) system. (d) The first passage time $\tau_{\rm fpt}^E$ versus the
gain/loss parameter $\gamma$ as it is measured from the energy exchange condition (see text). Open circles are the
experimental data while lines (of similar color) correspond to the theoretical prediction $\tau_{\rm fpt}^E=
\tau_{\rm fpt}/2$. The black dashed line denotes $\tau_{\rm fpt}^E$ for $\gamma=0$.
}
\label{fig:fig4}
\end{figure*}

The theoretical analysis (see reference \cite{SM}) relies on a Liouvillian formulation of Eqs.~(\ref{kirchhoff2})
which take the form
\begin{equation}
\label{liuvilian1}
{d\Psi\over d\tau} = {\cal L}  \Psi;\,
 {\cal L} = \left ( \begin{array}{cccc}
  0 & 0 & 1 & 0 \\
  0 & 0 & 0 & 1 \\
 -\alpha & \mu \alpha & \gamma & 0 \\
 \mu \alpha & -\alpha & 0 & -\gamma
 \end{array} \right)
\end{equation}
where $\Psi\equiv (Q_1, Q_2, {\dot Q_1}, {\dot Q_2})^T$.
Eq. (\ref{liuvilian1}) can be interpreted as a Schr\"odinger equation with non-Hermitian effective Hamiltonian $H_{\rm eff}=
i{\cal L}$. This Hamiltonian is symmetric with respect to generalized ${\cal P}_0{\cal T}_0$ transformations, i.e.
$[{\cal P}_0{\cal T}_0,H_{\rm eff}]=0$, where
\begin{equation}
\label{generaltransf}
{\cal P}_0=\left ( \begin{array}{cc}
\sigma_x & 0\\
0        &\sigma_x
 \end{array} \right);\quad
{\cal T}_0=\left ( \begin{array}{cc}
{\bf 1} & 0\\
0        &-{\bf 1}
 \end{array} \right) {\cal K}
\end{equation}
and $\sigma_x$ is the Pauli matrix, ${\bf 1}$ is the $2\times 2$ identity matrix, and ${\cal K}$ denotes the operation
of complex conjugation. By a similarity transformation ${\cal R}$,
\begin{equation}
{\cal R}=\left(
     \begin{array}{cccc}
       b+c & b+c & i & -i \\
       b-c & -(b-c) & i & i \\
       -(b-c) & b-c & i & i \\
       b+c & b+c & -i & i \\
     \end{array}
   \right)
  \label{rmatrix}
\end {equation}
$H_{\rm eff}$ can be related
to a transposition symmetric, $\cP\cT-$symmetric Hamiltonian $H$. Specifically,
\begin{eqnarray*}
H=H^T&=&{\cal R}H_{eff}{\cal R}^{-1},\qquad \cT=\cK={\cal R}\cT_0{\cal R}^{-1}\nonumber\\
\left[\cP\cT,H\right]&=&0,\quad \cP=\left(
                           \begin{array}{cccc}
                             0 & 0 & 0 & 1 \\
                             0 & 0 & 1 & 0 \\
                             0 & 1 & 0 & 0 \\
                             1 & 0 & 0 & 0 \\
                           \end{array}
                         \right)
={\cal R}\cP_0{\cal R}^{-1}
\end{eqnarray*}
where
\begin{equation}
\label{liuvilian2}
H =
\left ( \begin{array}{cccc}
  0 & b+i\gamma/2 & c+i\gamma/2 & 0 \\
  b+i\gamma/2 & 0 & 0 & c-i\gamma/2 \\
  c+i\gamma/2 & 0 & 0 & b-i\gamma/2 \\
  0 & c-i\gamma2 & b-i\gamma/2 & 0
 \end{array} \right)
\end{equation}
and
$b=\sqrt{(\a+\a^{1/2})/2}$, $c=-\sqrt{(\a-\a^{1/2})/2}$. This allows us to make contact
with the brachistochrone studies of Refs.~\cite{BBJM07,M07,GS08a,GS08b}.

The eigenfrequencies $\omega_{1,2}$ of system (\ref{kirchhoff2}) (or equivalently Eq. (\ref{liuvilian1})) are shown as
functions of the gain/loss parameter $\gamma$, in Fig.~\ref{fig:fig3} (solid lines). For $\gamma<\gamma_{\cal PT}=1/
\sqrt{1-\mu} - 1/\sqrt{1+\mu}$ the system is in the exact phase and thus the eigenfrequencies are real \cite{SLZEK11}.
We investigate the signal/energy tachistochrone passage under the constraint of fixed bandwidth $\delta \omega=\omega_1
-\omega_2$. Experimentally, the $\delta \omega$ constraint is implemented in the $LRC$ dimer through
adjustment of the mutual inductance.

Figure \ref{fig:fig3} shows both theoretical and experimental results for the parametric evolution of the eigenfrequencies
in the exact phase, for various $\mu$ values. The black dashed lines in Fig. \ref{fig:fig3} illustrate a path for
fixed $\delta\omega /\omega_0 = 0.36$ through the family of eigenfrequencies associated with different mutual inductances $\mu$.


Eqn.~(\ref{liuvilian1}) can be solved either analytically or via direct numerical integration in order to obtain the
temporal behavior of the capacitor charge $Q_n(\tau)$ and the displacement current $I_n(\tau)$ in each of the two
circuits of the ${\cal PT}$-symmetric dimer. For the investigation of the tachistochrone wave evolution, we consider an initial
displacement current in one of the circuits with all other dynamical variables zero. The first passage time
$\tau_{\rm fpt}$ is then defined as the time interval needed to reach an orthogonal state. In our experiments
this corresponds to the condition that the envelope function of the current at the initially excited circuit is
zero. We find that the first passage time is asymmetric with respect to the initially excited circuit. Specifically
we have that
\begin{equation}
\label{tautheory}
\tau_{\rm fpt}={1\over \delta\omega}\left[\pi\pm\arccos\left({\delta\omega^2-\gamma^2\over\delta\omega^2+\gamma^2}\right)\right]
\end{equation}
(see reference \cite{SM}) where the $+$ sign corresponds to an initial condition starting from the gain side while the $-$ sign corresponds to
an initial condition starting from the lossy side. For $\gamma\gg \delta\omega$, Eq. (\ref{tautheory}) takes
the limiting values $\tau_{\rm fpt}\approx 2\pi/\delta\omega$ and $\tau_{\rm fpt}\approx 2/\gamma$ respectively. The
latter case indicates the possibility of transforming an initial state to an orthogonal final one, or in more practical
terms, transferring energy from one side to the other, in an {\it arbitrarily} short time interval. In the opposite limit
of $\gamma=0$,
we recover for both initial conditions the Anandan-Aharonov lower bound for the first passage time $\tau_{\rm
fpt}=\pi/\delta\omega$ \cite{AA90}. This is the time for which energy is transferred from the initial circuit to
its partner according to the constraint of the Bandwidth theorem.

Geometrically, one can understand the relation (\ref{tautheory}) in the following way: the time required for the
evolution between two states induced by a Hermitian Hamiltonian is proportional to the length of the shortest
geodesic connecting the two states in projective Hilbert space \cite{AA90}. Non-Hermitian ${\cal PT}$-symmetric
Hamiltonians in the exact ${\cal PT}$-symmetric domain can be similarity mapped to equivalent Hermitian Hamiltonians.
Under such a similarity mapping the corresponding projective Hilbert space undergoes a deformation obtaining
a nontrivial metric. This results in an effective contraction or dilation of the corresponding geodesic and with
it of the corresponding evolution time \cite{M07,GS08b,M09}.

In Fig.~\ref{fig:fig4} we present some typical measurements for the temporal behavior of displacement currents. In
subfigure~\ref{fig:fig4}a we show $\left|I_1(\tau)\right|$ for an initial condition corresponding to the case $I_1(0) =1$
with all other dynamical variables zero. The case where the initial current excitation is at the lossy side i.e.
$I_2(0)=1$, is shown for contrast in Fig. ~\ref{fig:fig4}b. In both cases, agreement between the experiment
(circles) and the simulations (lines) is observed. For comparison, we also report with black line
the temporal behavior of the displacement current for the case of a passive circuit (i.e. $\gamma=0$) with the same
$\delta\omega$-constraint. We observe that the orthogonal target state is reached faster (or slower) depending on
whether the initial excitation is applied to the lossy (or gain) side.

The above results can be verified in more cases by changing the inductive coupling $\mu$ and gain/loss
parameter $\gamma$, while keeping constant the frequency difference $\delta \omega=\omega_2-\omega_1$. A summary of our
measured $\tau_{\rm fpt}$ versus $\gamma$ is presented in Fig.~\ref{fig:fig4}c. The experimental data show
agreement with the theoretical prediction Eq.~(\ref{tautheory}).


A parallel analysis pertains directly to the study of the energy transport from one side to another. Using the same
initial conditions as above we investigate the temporal behavior of the energies
\begin{equation}
 E_n(\tau)=\frac{1}{2} \frac{Q_n^2}{C} + \frac{1}{2}L I_n^2
\end{equation}
of each $n=1,2$ circuit. The first passage time can be defined as the time for which the two energies become equal for
the first time i.e. $E_1(\tau_{\rm fpt}^E)=E_2(\tau_{\rm fpt}^E)$. For passive (i.e. $\gamma=0$) coupled circuitry,
this time is half of the beating time $\tau_{\rm fpt}^E=\tau_{\rm fpt}(\gamma=0)/2$ and it is insensitive to the initial preparation.
In contrast, for the active ${\cal PT}$-symmetric dimer of Fig.~\ref{fig:fig2}, we find that the energy transfer from
the lossy (gain) side to the gain (lossy) one, is faster (slower) than the corresponding passive system with the same
$\delta\omega$. In Fig.~\ref{fig:fig4}d, we summarize our measurements for the $\tau_{\rm fpt}^E$ versus $\gamma$
under the constraint of fixed frequency bandwidth $\delta\omega$. A similar behavior as the one found for
the displacement current is evident.


Our results open a new direction towards investigating novel phenomena and functionalities of ${\cal PT}$-symmetric
arrangements with pseudo-unitary {\it spatio-temporal} evolution. Along these lines, we envision ${\cal PT}$-symmetric (nano)-antenna
configurations and metamaterial or optical microresonator arrays with unidirectional ultra-fast communication capabilities.
These structures also have potential applications as delay lines, buffers and switches. Questions like the effects
of non-linearity or the topological complexity of the ${\cal PT}$-symmetric structures in the tachistochrone dynamics
are open and offer new exciting opportunities, yet to be discovered.

{\it Acknowledgments}
We thank D. Christodoulides, D. Cohen, L. H\"uwel, V. Kovanis, and T. Morgan for useful discussions. This research was
supported by an AFOSR No. FA 9550-10-1-0433 grant, by an NSF ECCS-1128571 grant and by a Wesleyan Project grant.

\section{Supplemental Material}
\numberwithin{equation}{section}
\setcounter{equation}{0}
\renewcommand{\theequation}{A-\arabic{equation}}

From the eigenvalue equation $(H_{eff}-\om_k)\Xi_k=0$ with $k=1,2,3,4$ and the 
corresponding characteristic polynomial $\om_k^4-(2\a-\g^2)\om_k^2+\a=0$ one obtains $\om_{1,4}=\pm\sqrt{\Omega_+}$, 
$\om_{2,3}=\pm\sqrt{\Omega_-}$ where
\begin{eqnarray}
\Omega_\pm  &:= & \alpha-\frac{\gamma^2}2\pm \sqrt{\left(\alpha-\frac{\gamma^2}2\right)^2-\alpha}\nonumber\\
\Xi_k       & = & a_k(e^{-i\phi_k},e^{i\phi_k},-i\om_k e^{-i\phi_k},-i\om_k e^{i\phi_k})^T\nonumber\\
e^{2i\phi_k}&:= & \frac{\a-\om_k^2+i\g\om_k}{\a\mu},\quad a_k\in\RR.
\end{eqnarray}
For gain/loss parameters $\g\in[0,\g_{\cP\cT}]$ the eigenvalues are purely real, $\om_k\in\RR$; it holds $\phi_k\in\RR$ 
so that $\cP_0\cT_0\Xi_k=\Xi_k$ and the $\cP_0\cT_0-$symmetry is exact. For $\g>\g_{\cP\cT}$ the eigenvalues $\om_k$ are 
not real, but pairwise complex conjugate; one finds $\phi_k\not\in\RR$ so that $\cP_0\cT_0\Xi_k\not\propto \Xi_k$ and 
the $\cP_0\cT_0-$symmetry is spontaneously broken with $\cP_0\cT_0$ phase transition at $\g=\g_{\cP\cT}$. The Hamiltonian 
$H_{\rm eff}$ can be brought into the more symmetric form Eq. (5) via a similarity transformation 
${\cal R}$ given by equation (4). 
To describe the dynamics in the sector of exact $\cP_0\cT_0$ symmetry, $\g\in[0,\g_{\cP\cT}]$, where $\om_1=-\om_4$, $\om_2=-\om_3$, 
$\om_k\in\RR$ we start from an ansatz $\Psi(\tau)=\sum_{k=1}^4 e^{-i\om_k\tau}A_k\Xi_k$, $A_k\in\CC$ and impose the experimentally 
required reality constraint $\Psi(\tau)\in\RR^4$ as $A_1\Xi_1=\bar A_4\bar \Xi_4$, $A_2\Xi_2=\bar A_3\bar \Xi_3$. The solutions of 
the evolution equation (1), (2) with initial condition on the gain side $\Psi(\tau=0)=(0,0,1,0)^T$ are
\begin{eqnarray}
&&\Psi(\tau)=\frac{\a\mu}{\Delta}\left(
                                              \begin{array}{c}
                                                -\frac{\sin(\om_1 \tau+\d_1)}{\om_1}+\frac{\sin(\om_2 \tau+\d_2)}{\om_2} \\
                                                -\frac{\sin(\om_1 \tau)}{\om_1}+\frac{\sin(\om_2 \tau)}{\om_2} \\
                                                -\cos(\om_1 \tau+\d_1)+\cos(\om_2 \tau+\d_2) \\
                                                -\cos(\om_1 \tau)+\cos(\om_2 \tau) \\
                                              \end{array}
                                            \right)\nonumber\\
&&\nonumber\\
&&\Delta:=\om_1^2-\om_2^2\,,\qquad \frac{\a\mu}{\Delta}=\frac12\sqrt{1+\frac{\g^2}{\d\om^2}}\sqrt{1+\frac{\g^2}{\bar\om^2}}\nonumber\\
&&\d\om:=\om_1-\om_2,\qquad \bar\om:=\om_1+\om_2\nonumber\\
&&\sin(\d_1)=\frac{\g\om_1}{\a\mu}\,,\qquad
\cos(\d_1)=-\frac{\Delta-\g^2}{2\a\mu}\nonumber\\
&&\sin(\d_2)=\frac{\g\om_2}{\a\mu}\,,\qquad
\cos(\d_2)=\frac{\Delta+\g^2}{2\a\mu}\,.
\label{solution}
\end{eqnarray}
The first passage time $\tau_{\rm fpt}$ required for the evolution from the initial state $\Psi(\tau=0)=(0,0,1,0)^T$ to an orthogonal 
final state $\Psi(\tau=\tau_{\rm fpt})=(\psi_1,\psi_2,0,\psi_4)^T$, $\la\Psi(0)|\Psi(\tau_{\rm fpt}) \ra=0$ with regard to the slowly 
evolving enveloping amplitude follows from  $\psi_3(\tau)=\sqrt{1+\frac{\g^2}{\d\om^2}}\sqrt{1+\frac{\g^2}{\bar\om^2}}\,
\sin\left[\frac{\d\om\tau+\d_1-\d_2}2\right]\sin\left[\frac{\bar\om\tau+\d_1+\d_2}2\right]$ as
\begin{equation}
\label{time-2}
\tau_{\rm fpt}=(\d_2-\d_1)/\d\om.
\end{equation}
Due to the invariance of equations (1) under simultaneous action of $Q_1\rightleftarrows Q_2$ and $\g\to-\g$ the evolution from 
the lossy side $\Psi(\tau=0)=(0,0,0,1)^T$ to  $\Psi(\tau_{\rm fpt})=(\psi_1,\psi_2,\psi_3,0)^T$ requires a $\tau_{\rm fpt}$ obtainable 
from (\ref{solution}) by sign change $\g\to -\g$. Explicitly this yields  Eqs. (6).



\begin{thebibliography}{99}
\bibitem{note0} Note though Ref. \cite{M07} where these results have been questioned.
\bibitem{note} The tachistochrone passage, although closely related to
a brachistochrone evolution, may not necessarily reach the
brachistochrone limit for a corresponding non-Hermitian system.
\bibitem{note2}There are theoretical indications
for the existence of ultra-fast brachistochrone regimes in unstable dissipative setups as well \cite{assis-fring}.
\end{thebibliography}
\end{document}